\documentclass[twocolumn,showpacs,preprintnumbers,amsmath,amssymb]{revtex4}
\usepackage{graphicx}
\usepackage{dcolumn}
\usepackage{bm}
\usepackage{color}
\begin{document}
\preprint{APS/123-QED}
\title{
Novel Kondo-like behavior near magnetic instability in SmB$_6$
: temperature and pressure dependences of Sm valence
}
\author{N.~Emi$^1$}
\author{N.~Kawamura$^2$}
\author{M.~Mizumaki$^2$}
\author{T.~Koyama$^1$}
\author{N.~Ishimatsu$^3$}
\author{G.~Prist$\acute{\rm{a}}$$\breve{\rm{s}}$$^{4}$}
\author{T.~Kagayama$^5$}
\author{K.~Shimizu$^5$}
\author{Y.~Osanai$^6$}
\author{F.~Iga$^{6,7}$}
\author{T.~Mito$^1$}
 \email{mito@sci.u-hyogo.ac.jp}
\affiliation{
$^1$Graduate School of Material Science, University of Hyogo, Ako, Hyogo 678-1297, Japan\\
$^2$Japan Synchrotron Radiation Research Institute (JASRI), SPring-8, Sayo, Hyogo 679-5198, Japan\\
$^3$Graduate School of Science, Hiroshima University, Higashi-Hiroshima, Hiroshima 739-8526, Japan\\
$^4$Institute of Experimental Physics, Slovak Academy of Science, 04001 Ko$\breve{\rm{s}}$ice, Slovakia\\
$^5$KYOKUGEN, Graduate School of Engineering Science, Osaka University,
Toyonaka, Osaka 560-8531, Japan\\
$^6$College of Science, Ibaraki University, Mito Ibaraki 310-8512, Japan\\
$^7$Graduate School of Science and Engineering, Ibaraki University, Mito, Ibaraki 310-8512, Japan
}
\date{\today}

\begin{abstract}
We report a systematic study of Sm valence in the prototypical intermediate valence compound SmB$_6$.
Sm mean valence, $v_{\rm Sm}$, was measured by X-ray absorption spectroscopy as functions of pressure ($1<P<13$ GPa) and temperature ($3<T<300$ K).
Pressure induced magnetic order (MO) was detected above $P_c = 10$ GPa by resistivity measurements.
A shift toward localized $4f$ state with increasing $P$ and/or $T$ is evident from an increase in $v_{\rm Sm}$.
However $v_{\rm Sm}$ at $P_c$ is anomalously far below 3, which differs from the general case of nonmagnetic-magnetic transition in Yb and Ce compounds.
From the $T$ dependence of $v_{\rm Sm}(P,T)$, we found that $v_{\rm Sm}(P,T)$ consists of two different characteristic components:
one is associated with  low-energy electronic correlations
involving Kondo like behavior,
and the other with high-energy valence fluctuations.
\end{abstract}

\maketitle


The emergence of a wide variety of physical phenomena in $f$ electron systems, namely lanthanide and actinide compounds, is still of considerable interest.
One of key parameters governing the properties of the $f$ electron systems is hybridization between conduction and $f$ electrons ($c$-$f$ hybridization).
In some compounds, the effective $c$-$f$ hybridization is changeable depending on external parameters, including temperature $T$ and pressure $P$ \cite{PressureEffect}.
In a strongly hybridized state, a nonmagnetic ground state evolves and valence fluctuations of the lanthanide and actinide ions are brought about.
On the other hand, well localized $f$ electrons tend to form a long-range magnetic order (MO) through the Ruderman-Kittel-Kasuya-Yosida (RKKY) interaction.
These two characters generally compete with each other.
However, it is well known that some actinide compounds, particularly U compounds with multi $5f$ electrons, exhibit the duality between the itinerant and localized nature of $5f$ electrons.

The measurement of the valence of lanthanide ions is one of the most informative methods to evaluate
not only total angular momentum $J$, but also the degree of the localization:
in the case of Ce, Sm and Yb compounds, the trivalent state of the lanthanide ions corresponds to the strong localization of the $f$ electrons, while an increase in a divalent component (a tetravalent component for Ce) indicates a stronger hybridization.
Note that the application of this process to the U compounds is not easy because of difficulties in determining U valence experimentally.

In this letter, we propose that SmB$_6$ is one of a new class of lanthanide compounds showing a sort of duality from a systematic valence measurement.
SmB$_6$ is the prototypical intermediate valence compound with Sm mean valence of $\sim 2.6$ at room temperature and ambient pressure.
SmB$_6$ shows a semiconducting property with a narrow gap of $50 \sim 100$ K \cite{Kasuya,Takigawa}, whose origin is intimately related to the effectively $T$ dependent $c$-$f$ hybridization \cite{Yanase}.
The ground state of SmB$_6$ drastically changes with pressure:
the insulating gap collapses at $P_c =10$ GPa \cite{Derr} and simultaneously an MO phase appears below $\sim 12$ K \cite{Barla}.
However the details of $f$ state around $P_c$ has not been clarified despite intensive studies of this compound since 1960s.

For SmB$_6$, the estimation of the Sm valence at high pressures was attempted by various methods \cite{Ogita,NishiyamaJPSJ}.
A recent study of resonant x-ray emission spectroscopy indicates that the intermediate valent state persists up to at least 35~GPa and the monotonic $P$ dependence of the Sm mean valence, $v_{\rm Sm}$, is insensitive to the onset of MO\cite{Butch}.
Since valence fluctuations arise from high-energy electron dynamics reflecting Coulomb interactions in narrow quasiparticle bandwidth, the valence measured by high-energy tools may not be sensitive to nonmagnetic-magnetic changes in the ground state predominantly arising from low-energy magnetic correlations.
As a way out of the situation, we have carried out a systematic measurement of $v_{\rm Sm}$ by X-ray absorption spectroscopy (XAS) measurements for $1<P<13$ GPa and $3<T<300$ K.
The present results, in particular the $T$ dependence of $v_{\rm Sm}$ at different pressures, enabled us to notice the coexistence of two characteristic valence components:
one is associated with low-energy electronic correlations which seem to control the ground state in this compound, and the other is related to valence fluctuations in higher-energy scheme.

\begin{figure}[t]
\includegraphics[width=0.6\linewidth]{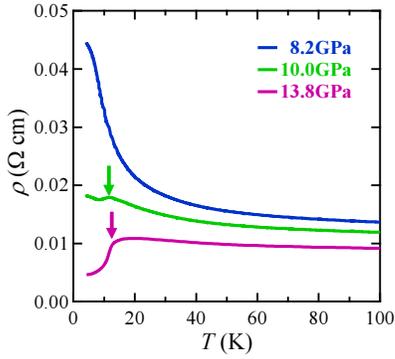}
\caption{\label{fig:epsart}
(color on line)
$T$ dependence of the resistivity $\rho$ at 8.2, 10.0, and 13.8 GPa.
The arrow indicates $T_{\rm M} \sim 12$ K.
Here, $T_{\rm M}$ is determined as the temperature at which $d^2 \rho / d T^2$ shows a minimum.
For details, see Ref. \cite{Suppl}.
}
\end{figure}

Single crystalline samples of SmB$_6$ were grown by a floating-zone method using an image furnace with four xenon lamps \cite{Iga}.
The XAS measurements near the Sm $L_3$-edge (6.72 keV) were performed at the beamline BL39XU of SPring-8, Japan \cite{Kawamura}.
Since the XAS spectrum was slightly dependent on the thickness of sample, we used a piece of SmB$_6$ crystal having partly constant thickness less than 10 $\mu$m.
The XAS spectra were recorded in the transmission mode using ionization chambers.
For the high pressure measurements, the sample was loaded in a diamond anvil cell (DAC) filled with a mixture of $4:1$ ${\rm methanol} : {\rm ethanol}$ as a pressure-transmitting medium.
Nanopolycrystalline diamond anvils were used to avoid glitches in the XAS spectra \cite{Ishimatsu}.
All pressures were applied at room temperature, and the data were acquired as the pressure cell was heated from the lowest temperature $\sim 3$ K.
The calibration of pressure was performed at each temperature of the measurement using the fluorescence from ruby chips mounted with the sample inside the DAC.
We have also carried out the high pressure measurement of the resistivity in order to detect the MO above $P_c$ as done in Ref.~\cite{Derr}.
For the resistivity measurement, we used a single crystalline sample from the same batch with that for the XAS measurement, and the measurement was performed using the D.C. four-terminal method.
The high pressure was generated using a DAC with NaCl as the pressure-transmitting medium, and the pressure was calibrated at the lowest temperature of each measurement by the ruby fluorescence method.

Figure~1 shows the $T$ dependence of the electrical resistivity $\rho$ measured at different pressures around $P_c$.
For $P < 10$ GPa, $\rho (T)$ reveals a semiconducting increase upon cooling down to $\sim 7$ K, followed by a tendency of saturation at lower temperatures.
Although we cannot discuss the absolute value of $\rho$ in detail due to ambiguity in the measurement of sample size, it is obvious that the semiconducting behavior is weakened near $P_c$ and $\rho (T)$ above 10.0 GPa exhibits a drop at $\sim 12$ K \cite{Suppl}.
This phenomenon and the boundary temperature $T_{\rm M} \sim 12$ K and pressure $P_c \sim 10$ GPa are in good agreement with the previous report \cite{Derr}, indicating the appearance of the MO above $P_c$ in our sample as well.

\begin{figure}[t]
\includegraphics[width=0.95\linewidth]{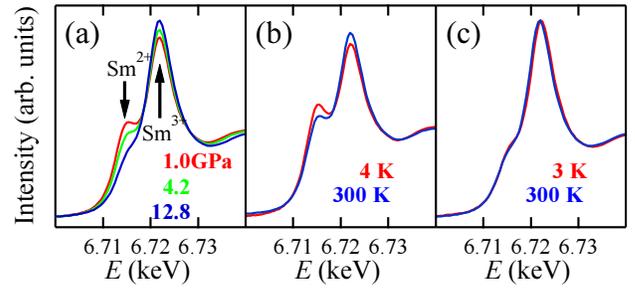}
\caption{\label{fig:epsart}
(color on line)
(a) $P$ dependence of Sm $L_3$-edge absorption spectra at 300 K.
(b), (c) $T$ dependence of XAS spectra measured at almost constant pressures (b) $P=0.6$ GPa (4 K) and 0.8 GPa (300 K), and (c) $P=12.4$ GPa (3 K) and 12.8 GPa (300 K)).
}
\end{figure}

Figures~2(a)-(c) show the Sm $L_3$-edge absorption spectra of SmB$_6$.
The intensity of the main peak at 6.722 keV and the shoulder like structure at 6.715 keV correspond to Sm$^{3+}$ and Sm$^{2+}$ components, respectively.
When pressure is increased at 300~K, the Sm$^{3+}$ component is enhanced, while the Sm$^{2+}$ component is reduced, consistent with previous reports \cite{Mizumaki,Butch} (see Fig.~2(a)).
As temperature increases, a similar shift toward the trivalent state appears at low pressures (Fig.~2(b)), however it becomes less pronounced under high pressure (Fig.~2(c)).
The observation of the two components is attributed to valence fluctuation at a time scale slower than that probed by XAS \cite{Suppl}.
$v_{\rm Sm}$ is estimated from the relative intensities of the Sm$^{2+}$ and Sm$^{3+}$ components in the XAS spectra.
Each component was modeled by the sum of Lorentz functions and an arctangent function representing the continuum excitations \cite{Suppl}.

The evaluated $v_{\rm Sm}$ is illustrated in Fig.~3(a) as functions of pressure and temperature.
The data points actually measured are also marked with the crosses.
The clamped pressure at room temperature inevitably varies with temperature (in most cases, the pressure decreases upon heating).
Therefore, we assumed a linear relation in between the neighboring $v_{\rm Sm}$ data points to construct this contour plot.
For $T<70$ K, we can regard the pressures to be almost constant as shown in Fig.~3(a).
$v_{\rm Sm}$ estimated in this $T$-range is presented in Fig.~3(b).

\begin{figure}[t]
\includegraphics[width=0.95\linewidth]{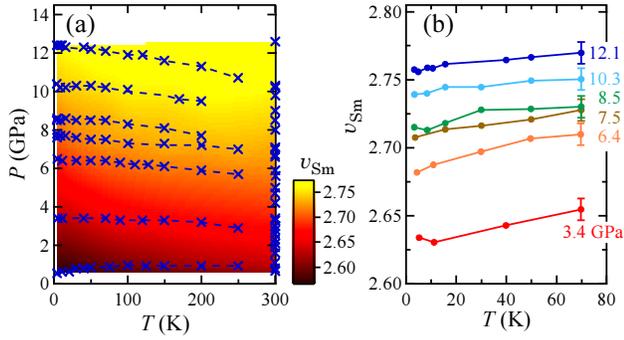}
\caption{\label{fig:epsart}
(color on line)
(a) $P$ and $T$ dependences of
$v_{\rm Sm}$.
The series of crosses connected by the broken lines are data points actually measured.
(b) $T$ dependence of $v_{\rm Sm}$ below 70 K at different pressures from 3.4 to 12.1 GPa.
The clamped pressures are regarded as being almost $T$ independent in this low temperature region.
}
\end{figure}

We extracted vertical $v_{\rm Sm}$-$T$ planes at constant pressure from Fig.~3(a), which is presented in the inset of Fig.~4(a).
As temperature increases at 1 GPa, $v_{\rm Sm}$ increases with a slight downward curvature, and the $v_{\rm Sm}$-$T$ curves are more monotonic and smoother over the whole $T$ range than shown in a previous report \cite{Mizumaki}.
With increasing pressure, $v_{\rm Sm}$ in the relatively higher temperature region ({\it e.g.} $T > 150$ K at 6 GPa) becomes less $T$ dependent.
The $T$ independent region gradually expands toward lower temperatures with pressure.
Such a pressure induced change in $v_{\rm Sm}$ is clearly demonstrated in Fig.~4(a), where we plot $\delta v'_{\rm Sm}(T) = v_{\rm Sm}(T) - v_{\rm Sm}(300{\rm K})$ at various pressures.
Although $\delta v'_{\rm Sm}(T)$ at 12 GPa still shows a small decrease below $\sim 50$ K, $|\delta v'_{\rm Sm} (4{\rm K})|$ is reduced by about 70 $\%$ compared to that at 1 GPa.
Therefore one expects that $|\delta v'_{\rm Sm}(4{\rm K})|$ will be further suppressed at higher pressures than 12 GPa.
In Fig.~4(b), we plot $-\delta v'_{\rm Sm}(4{\rm K})$ as a function of pressure.
$\delta v'_{\rm Sm}(4{\rm K})$ is extrapolated to zero at $P^{\ast} \sim 15$ GPa, indicating that $v_{\rm Sm}$ hardly depends on temperature above $P^{\ast}$.

The $P$ dependence of $v_{\rm Sm}$ at 300~K, $v_{\rm Sm}^{300}(P)$, is shown in Fig.~4(b).
Data at different temperatures are presented in Fig.~7(a) in Ref.~\cite{Suppl}.
As pressure increases, $v_{\rm Sm}$ increases but the slope $d v_{\rm Sm}/d P$ gradually decreases, consistent with the previous report \cite{Butch}.
Similar phenomenon, except for the large deviation of $v_{\rm Sm}$ from 3 at $P_c$, is also observed in several Yb heavy fermions which show the pressure induced nonmagnetic-magnetic transition (see Ref.~\cite{Suppl} or for example Refs.~\cite{Yamaoka1,Panella,Yamaoka2}).
In YbRh$_2$Si$_2$, which shows an antiferromagentic order at extremely low temperature of 70~mK and hence is thought to locate in the vicinity of nonmagnetic-magnetic criticality at ambient pressure \cite{Trovarelli,Gegenwart}, the Yb valence, $v_{\rm Yb}$, is larger than 2.9 as well \cite{Nakai}.
The proximity of $v_{\rm Yb}$ to 3 can be ascribed to the strong localization of $4f$ electrons due to marked lanthanide contraction characteristic of the Yb systems.
For Ce systems, Ce valence, $v_{\rm Ce}$, tends to deviate from 3 ($4f^1$) to 4 ($4f^0$) with pressure and the lanthanide contraction should be smaller.
CePd$_2$Si$_2$ and CeCu$_2$Ge$_2$ are driven through the magnetic-superconducting transition by pressure, and their $v_{\rm Ce}$ remains from 3.0 to 3.05 over the superconducting region (see Ref.~\cite{Suppl} or Refs.~\cite{YamaokaCePd2Si2,YamaokaCeCu2Ge2}).
Although there are only a few reports of $v_{\rm Ce}$ under pressure so far, we have found no serious exception in the Ce and Yb compounds in terms of the proximity to the trivalent state at the onset of the MO.

\begin{figure}[t]
\includegraphics[width=0.95\linewidth]{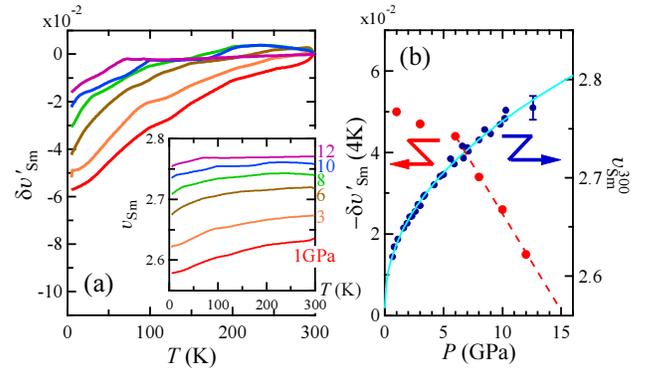}
\caption{\label{fig:epsart}
(color on line)
(a) $T$ dependence of $\delta v'_{\rm Sm}(T) = v_{\rm Sm}(T) - v_{\rm Sm}(300{\rm K})$ evaluated from Fig.~3(a).
Inset: $v_{\rm Sm}$-$T$ curves at representative constant pressures obtained from Fig.~3(b).
(b) The plots of $\delta v'_{\rm Sm}(4{\rm K})$ (left axis) and $v_{\rm Sm}^{300}(P)$ (right axis) as a function of pressure.
The broken line is a linear fit to the data for $P \geq 6$ GPa.
The solid line represents a fit $v_{\rm Sm}^{300}(P) = A P^{\alpha} + v_{\rm Sm}^{300}(0)$ to the data, giving $\alpha =0.47$ and $v_{\rm Sm}^{300}(0)=2.57$.
}
\end{figure}

The large deviation of $v_{\rm Sm}$ from 3 near the magnetic instability, which is also indicated in Ref.~\cite{Butch}, is observed in other Sm compounds as well.
SmS exhibits the pressure induced MO from the intermediated valence state (the so called ``gold phase'') at $P_c \sim 2$ GPa \cite{BarlaSmS}, and $v_{\rm Sm}$ at $P_c$ is comparable to that of SmB$_6$ \cite{Deen,Annese}.
SmOs$_{4}$Sb$_{12}$, which is the weak ferromagnet with $T_{\rm C}=3$ K, also shows $v_{\rm Sm} \sim 2.8$ \cite{MizumakiSmOs4Sb12}.
Interestingly, some of Eu compounds also show the MO away from the magnetic divalent state: for instance EuCu$_2$(Si$_x$Ge$_{1-x}$)$_2$ with $0<x<0.65$ shows an antiferromagnetic order at Eu valence $2.1<v_{\rm Eu}<2.4$ \cite{Fukuda}.
Therefore one may speculate that $4f^6$ state common to both Sm and Eu ions plays any role in assisting the long-range MO.
However considering contribution of its excited magnetic state (total angular momentum $J=1$) is unrealistic, because it should be extremely small due to a Boltzman factor involving excitation energy $\Delta$ (for instance, $\Delta=420$ K between the energy levels of $J=0$ and 1 for the divalent Sm ion \cite{Nickerson}).

According to the inset of Fig.~4(a), the $T$ dependence of $v_{\rm Sm}$ is characterized by two features:
(i) an almost $T$ independent term seen in the higher temperature and higher pressure region, and
(ii) a $T$ dependent part in the rest region.
For the latter, the $v_{\rm Sm}$-$T$ curve depends on pressure as well, as indicated by Fig.~4(a).
The deviation of $v_{\rm Sm}$ from the trivalent state is therefore described as the sum of the $T$ and $P$ dependent term $\delta v_{\rm Sm}(P,T)$ and the $T$ independent term $\Delta v_{\rm Sm} (P)$:
\begin{eqnarray}
v_{\rm Sm}(P,T) - 3 =  \Delta v_{\rm Sm}(P) + \delta v_{\rm Sm}(P,T).
\end{eqnarray}
Since $\delta v_{\rm Sm}(P,T)$ is easily suppressed by the thermal effects of a few hundred Kelvins, this term should be connected with the evolution of low-energy electronic correlations that can lead to the small modification of the Sm valence.
Such a $T$ dependence of $\delta v_{\rm Sm}$ is consistent with the $T$ dependent amplitude of $4f$ states in conduction bands predicted by recent study of dynamical mean field theory combined with density functional theory that was made for ambient pressure \cite{Denlinger}.
Although $\delta v_{\rm Sm}$ approaches 0 at high temperatures, $v_{\rm Sm} -3 < -0.2$, responsible for $\Delta v_{\rm Sm}$.
This term, hardly dependent on temperature up to at least 300~K, is related to the valence fluctuations in higher-energy scheme through the large $c$-$f$ hybridization.
The existence of the large hybridization may be consistent with predictions by the band structure calculations \cite{Yanase,Antonov}.
In the above-mentioned Yb and Ce compounds, the absolute value of the first term in Eq.~(1) is regarded much smaller than $\Delta v_{\rm Sm}$, implying that the valence fluctuations are weak.

\begin{figure}[t]
\includegraphics[width=0.75\linewidth]{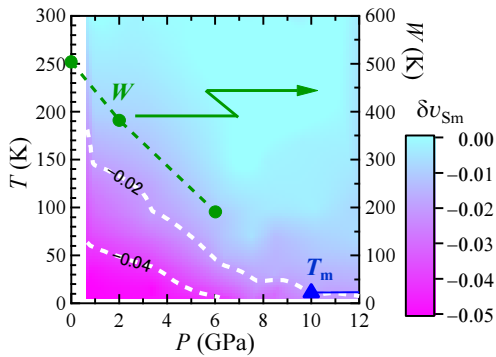}
\caption{\label{fig:epsart}
(color on line)
$P$ and $T$ dependences of $\delta v_{\rm Sm}$.
The triangle and solid line indicate $T_{\rm M}$ obtained from the present resistivity measurement.
The circles are quasiparticle bandwidth estimated by NMR studies (the right axis) \cite{Nishiyama}.
}
\end{figure}

In order to evaluate $\delta v_{\rm Sm}(P,T)$, we use the relation $\Delta v_{\rm Sm} (P) \cong v_{\rm Sm}^{300}(P) - 3$.
$v_{\rm Sm}^{300}(P)$ is consistent with the recent report \cite{Butch}.
Since $v_{\rm Sm}^{300}(P)$ approximately follows a function of $P^{\alpha}$ with $\alpha \sim 0.5$ \cite{Suppl}, we fit a power-law form $v_{\rm Sm}^{300}(P) = A P^{\alpha} + v_{\rm Sm}^{300}(0)$ to the data, giving $\alpha =0.47$ and $v_{\rm Sm}^{300}(0)=2.57$ (the solid line in Fig.~4(b) (right axis)).
$v_{\rm Sm}^{300}(P)$ does not saturate even at $P^{\ast}$ \cite{Butch}, where $\delta v_{\rm Sm} \sim 0$, suggesting the first and second terms in Eq.~(1) originate from different mechanisms.
Extracted $\delta v_{\rm Sm}(P,T)$ using Eq.~(1), which is equivalent to $\delta v'_{\rm Sm}$ presented in Fig.~4(b), is shown in Fig.~5.
Apparently the low temperature state possessing finite $\delta v_{\rm Sm}$ is largely suppressed by $P = 7-10$~GPa, in good agreement with $P_c$ \cite{comment_Pc}.
Figure~5 suggests that, if the electronic system in SmB$_6$ is cooled with little $\delta v_{\rm Sm}$, which is the case for $P \geq 10$ GPa, it falls into the magnetically ordered ground state.
Therefore the evolution of low-energy electronic correlations seems to impedes the long-range MO.

Interestingly, although $\Delta v_{\rm Sm}$ gives a dominant contribution in Eq.~(1) ({\it i.e.} $|\Delta v_{\rm Sm}| > 0.2 \gg |\delta v_{\rm Sm}|$), it is small $\delta v_{\rm Sm}$ that plays a key role in determining the ground state in this compound.
To examine the origin of $\delta v_{\rm Sm}$, we compare it with the $P$ dependence of quasiparticle bandwidth $W$ estimated from the measurement of nuclear spin lattice relaxation rate $T_1^{-1}$ of nuclear magnetic resonance (NMR) \cite{Nishiyama} in Fig.~5.
The $T_1$ measurement in heavy fermion materials predominantly probes the properties of heavy quasiparticles, since $T_1^{-1}$ is approximately proportional to the square of the density of states near Fermi level.
$W$ significantly decreases with pressure and seemingly approaches to zero around $P_c$, indicating that applying pressure induces the localization of $f$ electrons and the evolution of heavy fermion state, as $W \sim k_{\rm B} T_{\rm K}$, where $T_{\rm K}$ is Kondo temperature.
The good agreement between the $P$ dependences of $\delta v_{\rm Sm}$ and $W$ (see for example the contour line of $\delta v_{\rm Sm}=-0.02$) suggests that the increase in $|\delta v_{\rm Sm}|$ reflects the evolution of Kondo effect.

Indeed, Fig.~5 resembles the general phase diagram expected for the heavy fermions where the ground state changes from the so called Kondo lattice state to an MO phase as the localization of $f$ electrons is increased by an external parameter, including pressure.
Thus, Fig.~5 tells us that the finite $|\delta v_{\rm Sm}|$ and the MO are brought about by relatively localized electrons.
Recently, Barla {\it et al.} \cite{Barla} suggested that, in the low pressure region, $k_{\rm B} T_{\rm K}$ prevails over crystalline electric field splitting \cite{CEF} and magnetic moments cannot interact due to their uncorrelated fast fluctuations.
In this context, the suppression of $\delta v_{\rm Sm}$ at high pressures in Fig.~5 should be favorable for the appearance of long-range MO, because it is plausibly related to the decrease in $T_{\rm K}$ as mentioned above.

The phenomenologically extracted Eq.~(1) suggests the coexistence of two different characteristic valence components:
large $|\Delta v_{\rm Sm}|$ naturally indicates the existence of $f$ electrons considerably hybridized with conduction bands, while small $|\delta v_{\rm Sm}|$ is accounted for within Kondo regime for which relatively localized electrons are responsible.
To understand this, one may need to consider a peculiar mechanism such as the coexistence of weak and strong hybridizations in multi $f$ electrons system, as proposed in U compounds \cite{Sato}.
Yotsuhashi {\it et al.} have shown that the orbital dependence of hybridization and the effect of Hund's-rule coupling can bring about itinerant-localzied duality in a multi-localized electron system \cite{Yotsuhashi}.
On the other hand, the fact that $v_{\rm Ce}$ and $v_{\rm Yb}$ are close to 3 at the onset of MO in many Ce and Yb compounds suggests that distinct duality is hardly realized in the one $4f$ electron/hole configuration.
To detect the localized electronic character blurred in strongly hybridized electrons, further experiments in the vicinity of $P_c$, including NMR being sensitive to localized and/or magnetic moments, are indispensable.

In summary, we have carried out XAS measurements in the ranges of $1<P<13$ GPa and $3<T<300$ K to estimate the Sm valence of SmB$_6$.
The pressure induced MO of the sample was detected for $P > P_c =10$ GPa by resistivity measurements.
The increase in $v_{\rm Sm}$ with $P$ and/or $T$ indicates the localization of $4f$ electrons, analogous with the trend of Yb heavy fermion compounds which show a pressure induced nonmagnetic-magnetic transition.
However, the large deviation of $v_{\rm Sm}$ from 3 at $P_c$, as well as in some other Sm compounds, is remarkably different from the known cases in the Yb and Ce compounds.
We found that $v_{\rm Sm}$ is decomposed into two components:
one is associated with low-energy electronic correlations and the other with high-energy valence fluctuations.
The $P$-$T$ plot of the former component along with the MO phase closely resembles the general phase diagram of heavy fermions.

\begin{acknowledgments}
The authors are grateful to T.~Mutou, K.~Miyake, A.~Mitsuda, and H.~Harima for valuable discussions and H.~Sumiya and T.~Irifune for providing the nano-polycrystalline diamond anvils.
This work was supported by JSPS KAKENHI (Grants No. 24540349 and No. 16K05457).
The synchrotron radiation experiments were performed at the BL39XU of SPring-8 with the approval of the Japan Synchrotron Radiation Research Institute (JASRI) (Proposal Nos. 2014A1233, 2014B1564, and 2014B2041).
\end{acknowledgments}



\begin{thebibliography}{99}
\bibitem{comment_Pc}
The present and previous resistivity measurements \cite{Derr} indicate $P_c \sim 10$ GPa.
However, bulk specific heat measurements suggest that long range MO occurs for $P>8$ GPa \cite{Derr_J_Phys}.
\bibitem{CEF}
The resolution of the present XAS measurement is about $0.1 \sim 1$ eV, which is much larger than the realistic crystalline electric field (CEF) excitations of the order of $1 \sim 10$ meV. Therefore, even if effective CEF exists, we always count $f$ electrons in all the CEF-split levels, leading to a conclusion that the observed large deviation of $v_{\rm Sm}$ from 3 at $P_c$ is not ascribed to the CEF effect.
\end{thebibliography}

\end{document}